\documentclass[usenatbib]{mnras}

\usepackage{newtxtext,newtxmath}

\usepackage[T1]{fontenc}
\usepackage{ae,aecompl}

\usepackage{multirow,tabularx}
\usepackage{amsmath}
\usepackage{graphicx}
\usepackage{BibDef}
\usepackage{xcolor}
\usepackage[normalem]{ulem}
\usepackage{hyperref}
\hypersetup{
	colorlinks=true,        
	linkcolor=blue,         
	citecolor=blue,         
}

\newcommand{\msun}{ M$_\odot$ }

\newcommand{\lsim}{\mathrel{\rlap{\lower 3pt \hbox{$\sim$}} \raise 2.0pt \hbox{$<$}}}
\newcommand{\gsim}{\mathrel{\rlap{\lower 3pt \hbox{$\sim$}} \raise 2.0pt \hbox{$>$}}}
\newcommand{\Fstot}{$F_{\rm scat,tot}$}

\newcommand{\Fdir}{$F_{\rm dir}$}

\title[MBHB light curve polarization]{Binary black hole signatures in polarized light curves}

\author[M. Dotti et al.]{Massimo Dotti$^{1,2}$\thanks{E-mail: massimo.dotti@unimib.it},
Matteo Bonetti$^{1,2}$\thanks{E-mail: matteo.bonetti@unimib.it},
Daniel J. D'Orazio$^{3}$,
Zolt\'an Haiman$^{4}$,
\newauthor 
and Luis C. Ho$^{5,6}$
\\
$^{1}$Universit\`a degli Studi di Milano-Bicocca, Piazza della Scienza 3, 20126 Milano, Italy\\
$^{2}$INFN, Sezione di Milano-Bicocca, Piazza della Scienza 3, 20126 Milano, Italy \\
$^{3}$Niels Bohr International Academy, Niels Bohr Institute, Blegdamsvej 17, 2100 Copenhagen, Denmark\\
$^{4}$Department of Astronomy, Columbia University, New York, NY, 10027, USA\\
$^{5}$Kavli Institute for Astronomy and Astrophysics, Peking University, Beijing 100871, China\\
$^{6}$Department of Astronomy, School of Physics, Peking University, Beijing 100871, China
}

\date{Accepted XXX. Received YYY; in original form ZZZ}

\pubyear{2021}

\begin{document}
\label{firstpage}
\pagerange{\pageref{firstpage}--\pageref{lastpage}}
\maketitle

\begin{abstract}
  Variable active galactic nuclei showing periodic light curves have been proposed as massive black hole binary (MBHB) candidates. In such scenarios the periodicity can be due to relativistic Doppler-boosting of the emitted light. This hypothesis can be tested through the timing of scattered polarized light. Following the results of polarization studies in type I nuclei and of dynamical studies of MBHBs with circumbinary discs, we assume a coplanar equatorial scattering ring, whose elements contribute differently to the total polarized flux, due to different scattering angles, levels of Doppler boost, and line-of-sight time delays. We find that in the presence of a MBHB, both the degree of polarization and the polarization angle have periodic modulations. The minimum of the polarization degree approximately coincides with the  peak of the light curve, regardless of the scattering ring size. The polarization angle oscillates around the semi-minor axis of the projected MBHB orbital ellipse, with a frequency equal either to the binary's orbital frequency (for large scattering screen radii), or twice this value (for smaller scattering structures). These distinctive features can be used to probe the nature of periodic MBHB candidates and to compile catalogs of the most promising sub-pc MBHBs. The identification of such polarization features in gravitational-wave detected MBHBs would enormously increase the amount of physical information about the sources, allowing the measurement of the individual masses of the binary components, and the orientation of the line of nodes on the sky, even for monochromatic gravitational wave signals.
\end{abstract}


\section{Introduction}

Massive black hole (MBH) binaries (MBHB), i.e. pairs of MBHs
gravitationally bound to each other, have been predicted to form during the hierarchical growth of galaxies and to be observable if at least one component of the binary shows some level of accretion activity \citep[e.g.][]{BBR80}. A definite observational confirmation of any MBHB has not yet been found. The most promising MBHB candidate is hosted by the radio-galaxy 0402+379 \citep{Rodriguez09}, where two flat-spectrum radio cores have been detected through radio interferometry at a projected separation of $\approx 7$ pc. Depending on the separation between the two sources on the line of sight (l.o.s.) and on the (poorly constrained) total mass of the MBHs, this double radio source either represents the closest MBH pair known or the only genuine (i.e. gravitationally bound) MBHB imaged to date.

Because of the exceptional angular resolution required, no other MBHB candidates have been imaged so far \citep[e.g.][]{BurkeSpolaor11, DOrazioLoeb:2018}. Other MBHB signatures have been proposed and searched for. The most studied is the predicted presence of single or double broad emission lines (BELs) shifted with respect to the host galaxy rest frame, and drifting in time as a consequence of the orbit of the two MBHs around their centre of mass \citep{BBR80}. Such a signature has been thoroughly searched for in large spectroscopic data-sets either focusing on large spectral shifts between broad and narrow lines \citep{Tsalmantza11, Eracleous12} or on BELs centred at different frequencies at different epochs \citep{Ju13, Shen13, Wang17}. While some of the candidates have been definitely disproved through dedicated observational follow-ups \citep[see e.g. the case of SDSS J092712.65+294344.0,][]{Decarli14}, no spectroscopic candidate has emerged as a clear MBHB, and different scenarios that can explain their peculiar spectral features are available \citep[see, e.g.][]{DSD12}. Furthermore the presence of clearly shifted BELs is expected only in a limited range of binary separations, when the corresponding orbital period is typically $\gsim 10-100$ yr \citep{Montuori11, Montuori12, 2019ApJ...870...16N,2021MNRAS.500.4065K},
making the mapping of the whole MBHB evolution particularly challenging.

On the other hand, at separations $\lsim 0.01$ pc, smaller than those characteristic of spectroscopic binary candidates, many theoretical studies have predicted a significant variability in the observed nuclear light curve due to different physical processes. For example, in studies of the evolution of MBHBs in circumnuclear discs, a modulated gas inflow from the outer gas distribution to the mini-discs bound to each individual MBH is commonly observed
\citep{AL94,IPP99,HMH08, Cuadra09, Roedig11, Roedig12, DHM:2013:MNRAS, Farris15, D'Orazio:CBDTrans:2016, Tang17,Mirandaetal:2017,Bowen18, Dascoli18} as a consequence of the non-axisymmetric and time-dependent potential of the binary. Such modulated inflow could result in a similarly variable luminosity, depending on the properties of the in-flowing gaseous streams and of the pre-existing mini-discs \citep[see the discussion in][]{Sesana12}. An alternative cause of observed variability could be the plunging of a very eccentric secondary MBH onto the primary disc, as proposed by \cite{Valtonen08} for the observed variability of OJ287. Finally, even in the absence of periodic inflows or very eccentric binaries (as expected in the case of a low-mass secondary; \citealt{D'Orazio:CBDTrans:2016,Duffell+2020}), variability can be caused by the 
relativistic Doppler boost
of the emitted spectrum during the orbit of the MBHB, resulting in a variable flux observed in fixed observational bands, as proposed for PG 1302-102 in \cite{DHS15}. This last model has the peculiarity of predicting different variability amplitudes at different wavelengths, as demonstrated for the UV vs. optical light curves of PG 1302-102~\citep{Xin+2019}.

Periodicity analyses of observed quasar light curves, over multiple wavelengths, has led to a growing MBHB candidate list
\citep[e.g.][]{Ackermann15, Graham15, Li2016, Charisi16, Sandrinelli16,Sandrinelli18, Severgnini18, Li+2019, LiuGez+2019,Chen+2020}.\footnote{NGC 5548 is particularly interesting since, as discussed in \citet{Li2016}, it shows a periodic modulation of the broad H$_\beta$ line profile as well. For this object, the parameters of the binary that best describe the data imply that the continuum variation is not related to Doppler-boosting.}
Theoretically, a sizable fraction of such periodically modulated binaries is predicted, with a considerable fraction of them being caused by the pure Doppler-boosting process described above \citep{Kelley19}. Still, it has been suggested that the number of candidates observed in the Catalina Real-time Transient Survey \citep[CRTS,][]{Drake09} and in the Palomar Transient Factory \citep[PTF, ][]{Rau09} would imply a total population of MBHBs inconsistent with current upper limits to the gravitational wave (GW) background posed by pulsar timing array (PTA) campaigns \citep[unless the SMBH masses are systematically overestimated or if the typical binary mass ratios are small, see][]{Sesana18}. A similar result applies to Blazar candidates \citep{Holgado18}.\footnote{The current PTA upper limits on the GW background also limits to $\lsim 20 \%$ the fraction of MBHBs in ultra luminous infrared galaxies (ULIRGs) that are allowed to merge within a Hubble time \citep{Inayoshi18}.}

The rate at which variability-selected MBHB candidates are discovered will ramp-up significantly with forthcoming time-domain surveys \citep[see the discussion in][]{Kelley19}, and additional tests are clearly required in order to probe the true nature of these candidates. In this study we predict some peculiar features that
can be observed in the light curve of such candidates, when the
polarized light, scattered into our line of sight by a circumbinary torus, is considered. In the Doppler-boosting scenario a related
characteristic signature in the infrared light curve,``reverberating'' from a dusty circumbinary torus, has been
predicted by \cite{DH17}, but current data have not been able to
discriminate between such scenario and alternative models to date. Here we propose an alternative observational signature in the {\em polarized} light curve of periodically variable MBHB candidates. We investigate how, in the Doppler-boosting case, the interaction between the un-polarized light of an accretion disc (bound to one of the binary components) and a circumbinary scattering ring affects the polarization degree of the observed flux, and, similarly to the reverberation mapping case \citep{BM82}, how the polarized flux is shifted in time with respect to the direct flux observed.

This paper is organised as follows: in \S~\ref{sec:model} we describe the model assumed to compute the time evolution of the direct and the scattered light. Our results, with particular emphasis on the time evolution of the polarization fraction, are presented in \S~\ref{sec:results}. In \S~\ref{sec:discussion} we discuss some aspects of the observability of the predicted signatures. Finally, in \S~\ref{sec:conclusions} we present our conclusions.

\section{Modelling of the polarization variability}\label{sec:model}

\begin{figure*}
    \centering
    \includegraphics[scale=0.35]{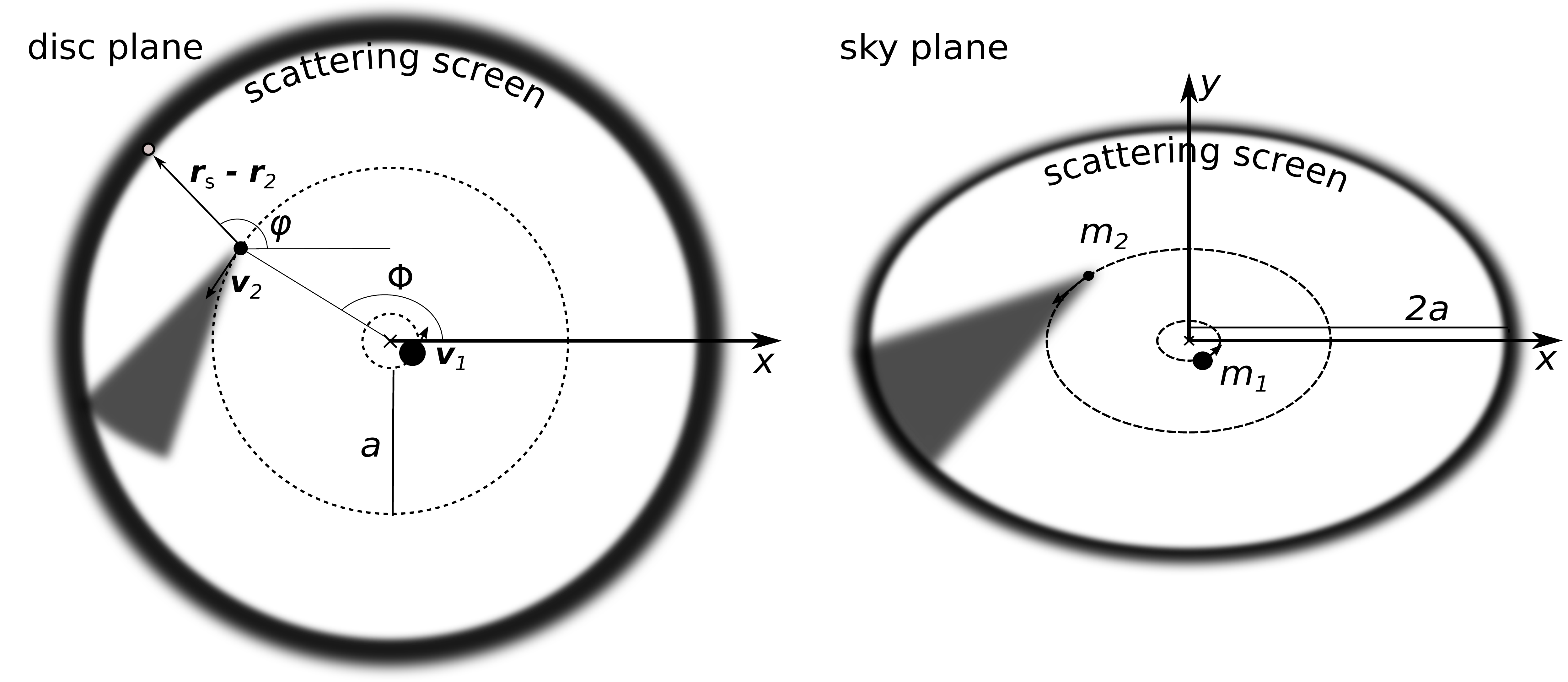}
    \caption{Sketch of the model for the binary-circumbinary system. The left and right panels refer to a face-on view and to the system as observed in the plane of the sky. In the latter view, the upper edge of the scattering screen (and of the binary orbit) is the closest to the observer, so that the secondary is approaching the observer for $\phi=0^{\circ}$. In the same panel the $x$ and $y$ axes described in the text are marked. The gray shaded area indicates the direction in which the accretion disc emission is maximally boosted in the reference frame of the binary centre of mass. The $x$ axis is defined to coincide with the line of nodes, which is the same in the two projections.} 
    \label{fig:sketch}
\end{figure*}

Here we describe a simple model used to characterize our new polarimetric test for binarity.
The model consists of a MBHB similar to the candidate described in \cite{DHS15}: total mass $M_{1+2}=2\times 10^9 \rm ~M_{\odot}$, rest-frame period $\tau \approx 4$ yr (corresponding to an observed period of 5.2 yr for redshift $z\approx 0.3$), a separation between the two MBH $a=0.015$ pc and mass ratio $q=M_2/M_1=0.1$. The binary is assumed circular. Only the secondary MBH is assumed to be accreting, with a luminosity that does not depend on time and is isotropic in the MBH reference frame.\footnote{Numerical simulations pointed out that the secondary can be significantly more luminous than the primary in the mass ratio regime considered here \citep[e.g.][]{Cuadra09, Roedig11,Farris15}. Here we assume the primary as inactive for simplicity.}

The binary is surrounded by a co-planar axi-symmetric circumbinary disc whose inner radius is located at $R_{\rm cbd}=2 a$, as expected for circular binaries \citep{AL94}. We consider the polarization to be due to Thomson scattering from free electrons orbiting in a thin disc co-planar with the accretion disc, as proposed by e.g. \cite{Antonucci84} and \cite{Smith02, Smith04} to justify the observational polarization properties of Type I AGN. In our reference model, we consider the scattering to happen in a narrow ring at the inner edge of the disc ($R_{\rm screen}=R_{\rm cbd}\approx 0.03$ pc), due to the large gas density accumulating there because of the torque exerted by the binary onto the disc \citep{LP79a, LP79b, AL94, SC95, GR00}.\footnote{This radius is broadly consistent with the
estimates used by, e.g. \cite{Smith02, Smith05} to model the spectropolarimetric properties of Seyfert I's, as well as with the
size of the scattering screen measured for NGC 4151 by
\citet{Gaskell12} through the reverberation of the polarized
continuum following the total flux variations.} Larger scattering screen sizes are considered as test cases, up to  $R_{\rm screen}\approx 6$ pc.  The
binary-circumbinary disc system is inclined by an angle $\theta$, defined as the angle between the binary angular momentum and the line-of sight (assumed to lie on the $z$ positive axis).

The direct apparent flux $F_{\nu, \rm dir}$ (i.e. the light that does not undergo any scattering before being detected) measured by the observer at a fixed frequency $\nu$ will change as a function of the secondary phase $\phi$, due to the combined effect of time dilation, light aberration and the blue/red-shift of the spectrum. To the first order in  $v_{\rm 2,Z}/c$ the modulation reads: 

\begin{equation}\label{eq:boost}
    \frac{\Delta F_{\rm dir}}{F_{\rm dir}}=(3-\alpha)\left(\frac{v_{\rm 2,Z}}{c}\right)
\end{equation}
where $v_{\rm 2,Z}=v_2 \cos(\phi) \sin(\theta)$ is the secondary velocity component along the l.o.s., $\phi$ is the azimuthal angle of the secondary measured in the binary's orbital plane, $\alpha$ is the exponent of the power-law that best describes the spectrum in the frequency region of interest \citep{DHS15} and $c$ is the speed of light. We follow the approximation from \cite{DHS15} and assume $\alpha =1.1$ as a good proxy for the optical V band.

The total observed scattered light (\Fstot) is the sum of the contributions from all scattering elements at the edge of the circumbinary disc, evaluated at the appropriate retarded time. It must be stressed that:
\begin{enumerate}
    \item {\bf as in a ``standard'' single central MBH scenario}, each scattering element contributes differently to the total flux due to the scattering geometry that can partially or totally suppress polarizations. The dependence of such an effect on the relative positions of the secondary and each screen element are detailed below;
    \item {\bf differently from the ``standard'' single MBH case}, the secondary does not lie at the centre of the scattering screen (i.e. the relative separation between each scattering element and the secondary
    ${\bf d}_{\rm 2-screen} = \mathbf{r}_s-\mathbf{r}_2$ 
    is not constant, see Fig.~\ref{fig:sketch}). Each scattering element is therefore irradiated with a flux modulated by a $(1/d_{\rm 2-screen})^2$ term;
    \item {\bf differently from the ``standard'' single MBH case}, the relative velocity between the secondary and each scattering element ${\bf v}_{\rm 2-screen}$ results in a Doppler boost as observed by each screen element, i.e., the Doppler boost described in equation~\ref{eq:boost}, where the projection of ${\bf v}_{\rm2-screen}$ on ${\bf d}_{\rm 2-screen}$ must replace $v_{\rm 2,Z}$;
    \item {\bf as in a ``standard'' single central MBH scenario}, we consider a second Doppler-boost due to the relative motion of the screen scattering element with respect to the observer.  
\end{enumerate}

The geometry of the system is sketched in Fig.~\ref{fig:sketch} for the $R_{\rm screen}=2 a$ case, where the direction in which the effect of the Doppler-boost is maximized is highlighted by the shading.

The contribution of each screen element to $F_{\rm scat,tot}$ is computed as follows. 
The radiation originally emitted by the minidisc and pointing toward a screen element (the $i^{\rm th}$ screen element in the following description) is assumed to be completely unpolarized. It can therefore be decomposed in two equally intense perpendicular linear polarizations. One of these two polarizations (P1$_{\rm i}$) is chosen to be perpendicular to the plane defined by the directions of propagation of light before and after the scattering with each screen element. The reason behind such choice is that P1$_{\rm i}$ is the only one whose flux is not reduced by the scattering onto the $i$-th screen element.
Being perpendicular to the direction of propagation of the light after the scattering, P1$_{\rm i}$ is forced to lie in the plane of the sky. The angle $\gamma$ between P1$_{\rm i}$ and the semi-major axis of the projected orbital ellipse (the $x$ axis according to Fig.~\ref{fig:sketch}) satisfies the following relation (which follows from a cross product):
\begin{equation}\label{eq.angpol}
\tan (\gamma) = -\frac{\sec (\theta)}{\tan (\varphi)},
\end{equation}
where $\varphi$ is the angle between ${\bf d}_{\rm 2-screen}$ and the $x$ axis for that specific screen element. 
The flux associated with the perpendicular polarization (P2$_{\rm i}$) is maximally reduced by the scattering, by a factor 
$\sin^2(\varphi) \sin^2(\theta)=\cos^2(\theta_{\rm scat})$, where $\theta_{\rm scat}$ is the scattering angle (i.e. the angle between the incoming light's direction and the line of sight). Therefore, the total flux scattered by a single $i$-th screen element is \citep{Rybicki1979}: 
\begin{equation}\label{eq:scatteredfulx}
F_{\rm scat, i} = F({\rm P1}_{i})+F({\rm P2}_{\rm i})=F_{\rm pre-scat,i} \left( \frac{1+\cos^2{\theta_{\rm scat}}}{2}\right),
\end{equation} 
where $F(P1_{\rm i})$ and $F(P2_{\rm i})$ are the scattered fluxes associated to the two polarizations after the scattering, while $F_{\rm pre-scat,i}$ is the flux incident onto the $i$-th scattering element, computed considering the modulations given by the varying $d_{\rm 2-screen}$ and relative velocity between the active MBH and the screen element (see above). 

The observed total scattered flux $F_{\rm scat}$ at a given time $t_{\rm obs}$ is then computed as the sum of all the contributions for each individual scattering element, where each contribution is evaluated taking in consideration the position and velocity of the secondary MBH at  the correct  $t_{\rm emit}$ time to take into consideration the different light travel times from the secondary to the scattering elements and then to the observer:\footnote{Since all the rays have to travel the distance
between the observer and the centre of mass of the binary, we compute
only the relative delay.}
\begin{equation}\label{eq:t1}
  \tau_{\rm scattered} = t_{\rm obs}-t_{\rm emit}= \frac{|{\bf d}_{\rm 2-screen}| -  z_{\rm screen}}{c},
\end{equation}
where $z_{\rm screen}$ is the $z$ component of the position of each screen element. A similar correction is considered for the direct flux:
\begin{equation}\label{eq:t2}
  \tau_{\rm direct} = -\frac{z_2}{c},
\end{equation}
where $z_2$ is the coordinate along the l.o.s. of the secondary. 
The total scattered flux (as well as all its polarizations) is normalized so that its average value over a MBHB orbit is equal to $f$ times the average value of $F_{\rm dir}$ over the same time span, where 
the constant $f<1$ parameterizes all of the un-modeled uncertainties (e.g. the optical depth and clumpiness of the screen) that do
not allow us to predict the actual intensity of the light scattered in
the $z$ direction.\footnote{We stress that the details subsumed by the factor $f$ are not important for the main purpose of this exercise, which is to show that the polarized scattered light has a characteristic light curve that differs from the direct light, allowing for a test of the binary model.}
The total observed flux ($F_{\rm tot}$) at any given time is the sum of the direct and scattered flux.

The polarization fraction of the observed flux and its polarization angle on the sky (defined below) are computed as follows. We compute how much flux would be observed when selecting only the polarization with an angle $\beta$ with respect to the $x$-axis (i.e. the line of nodes for the binary's orbital plane).
This would be the flux observed when a polarimetric filter orientated at an angle $\beta$ with respect to the line of nodes is applied to the observing instrument.  
The value of the flux is obtained by adding all the polarization components (both of the direct and scattered light) after projecting them onto the direction of the filter:
\begin{equation}
    F_{\beta}=\frac{1}{2}F_{\rm dir}+\sum_i \left[F(P1_{\rm i})\cos^2(\gamma-\beta)+F(P2_{\rm i})\sin^2(\gamma-\beta)\right], 
\end{equation}
where the first term on the right hand side of the equation does not depend on the orientation of the filter because the direct light is assumed unpolarized, while the last term is projected with the factor $\sin(\gamma-\beta)$ because the second polarization axis is perpendicular to the first one. We then numerically search for the value of $\beta$ that maximises $F_{\beta}$ (i.e. the polarization angle of the observed flux, $\beta_{F_{\rm max}}$ hereafter),\footnote{The value of $\beta$ minimizing the observed flux through the polarimetric filter differs by $\pi/2$ from the angle that maximises it, by construction.}
and compute the maximum and minimum values of $F_{\beta}$ ($F_{\rm \beta, max}$ and $F_{\rm \beta, min}$, respectively). The polarization fraction is then computed as:
\begin{equation}
    P=\frac{F_{\rm \beta, max}-F_{\rm \beta, min}}{F_{\rm tot}}.
\end{equation}

We are aware of the many simplifying assumptions made in the modeling
of the binary/scattering screen system. The screen is modeled as a
circular narrow ring of electrons whose covering factor is only
considered as a normalization of the scattered light. Any realistic
geometry of the ring will be extended, and the ring is expected to show
clear deviations from axisymmetry due to the time-dependent potential
of the MBHB \citep[e.g.][]{Shi12,DHM:2013:MNRAS}.
In general, we expect deviations from such
over-simplified assumptions that allow for the study of the effect of the
three dimensional structure of the screen would smear in time the
peculiar features of the scattered light for a single screen. 
In any case, it is clear that such smearing effect would, if anything, strengthen our main conclusion, i.e. that the polarization degree has a minimum in correspondence of the total flux maximum.

We refer the reader to \cite{SMP19}
for an independent study of the spectro-polarimetric
properties on MBHB broad lines. Those authors also assume a flattened
equatorial free-electron ring as the scattering screen, with a finite
radial extension between minimum and maximum radii of 0.1 and 0.5 pc respectively (similar to the largest
scattering screen considered here), and a half-opening angle of
30$^\circ$ with respect to the equatorial plane. Differently from
our simple model, \cite{SMP19} performed MonteCarlo radiative
transfer realizations for different broad line region geometries
using the STOKES code \citep{Goosmann07, Marin12, Marin15, Marin18,
RojasLobos18}. The main differences between their study and the present one
are: $(i)$ we focus on the polarization properties of the continuum,
while \cite{SMP19} focus on the broad emission line properties,
$(ii)$ \cite{SMP19} focus on the details observable in single polarized light spectra, while we focus on the time
evolution of the polarization properties, and $(iii)$ \cite{SMP19} do not consider the Doppler-boosting effect due to the motion of the secondary, which, as discussed in the following section, is the dominant effect in determining some of the observational features in our case. 

\section{Results}
\label{sec:results}

\begin{figure}
    \centering
    \includegraphics[width=0.48\textwidth]{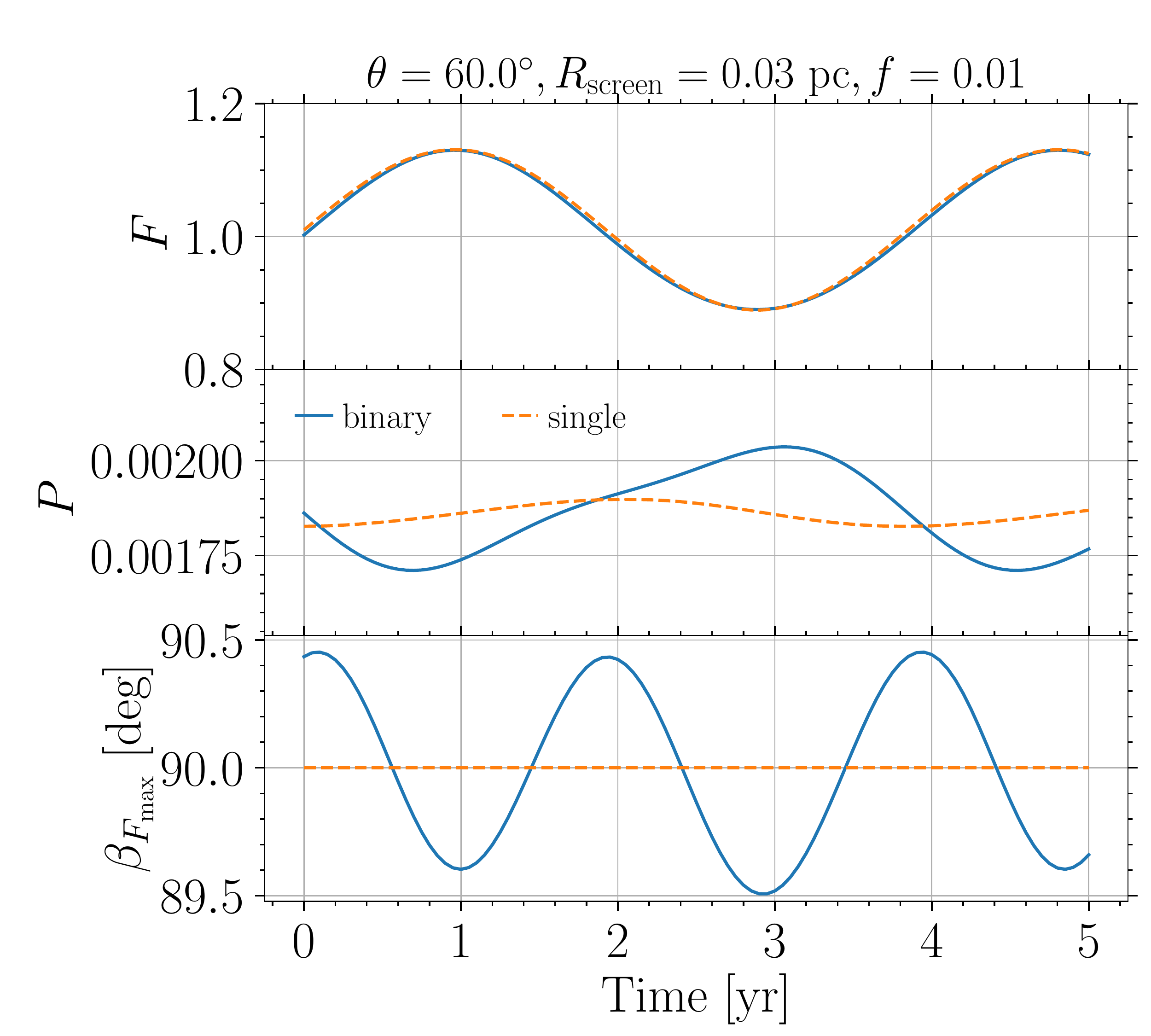}
    \caption{Blue line: mock light curve (upper panel), polarization fraction (middle panel) and polarization angle (lower panel) for the reference model, assuming $\theta=60^{\circ}$, $f=0.01$ and $R_{\rm screen}=2 a=0.03$ pc, corresponding to the inner edge of the circumbinary disc. The orange dashed lines show for comparison a test case with a single MBH at rest in the centre of the scattering ring emitting an isotropically pulsating light with the same properties as the observed \Fdir~ of the binary reference model.}
    \label{fig:res1}
\end{figure}

The blue solid lines in Figure~\ref{fig:res1} show the results of our binary model under the assumption of a fraction of scattered light $f=0.01$, an inclination $\theta=60^{\circ}$ and $R_{\rm screen} =2\,a=0.03$ pc. 
The upper panel refers to the time evolution of the total detected flux. A modulation is imparted by the binary motion on the direct light, and, due to the small contribution of scattered light to the total flux, the latter is modulated with the same period, as clearly observable in figure. 
The middle panel shows the evolution of the polarization fraction $P$, which varies on the same timescale, but which has its maximum close to the minimum of the direct flux (and, therefore, of the total flux), that, being unpolarized, suppresses $P$. The bottom panel shows the evolution of the polarization angle $\beta_{F_{\rm max}}$, which oscillates around a central value of 90$^\circ$. Such value is due to $(i)$ the specific orientation of the reference frame chosen (with the $x$-axis parallel to the semi-major axis of the projected orbital ellipse), and $(ii)$ the fact that the the scattered light is more polarized when the scattering angle is closer to $90^\circ$~(eq.~\ref{eq:scatteredfulx}). As an example, when the secondary is crossing the $x$-axis, the flux scattered by the screen elements with $\varphi=0 ~ (\pi)$ is maximally polarized, and the only surviving polarization has exactly $\beta_{F_{\rm max}}=90^{\circ}$. The frequency of the oscillations of $\beta_{F_{\rm max}}$ is twice that of the MBHB orbit, due to the symmetries in our model (circular MBH orbit and circular scattering screen), and to the negligible smearing effect of the time delays for the small radii assumed for the scattering screen. 

We checked whether such behaviour is distinctive of the MBHB scenario or if it is expected more generically, when modulations of the continuum are present, by computing the polarization degree in the case of a single "pulsating" MBH at rest at the centre of the scattering screen (orange-dashed lines in figure~\ref{fig:res1}). 
This test case replicates the Doppler-modulation, except the pulsations are set to be isotropic, allowing us to assess the importance of the anisotropically beamed nature of the Doppler boost. 
In this case the only Doppler-boost is due to the motion of the scattering screen elements with respect to the line of sight. The same sinusoidal evolution of \Fdir~  results in a lower polarization degree $P$ showing a minimum (maximum) closer to the points of maximum slope of the direct light with respect to the binary case.
More importantly, the polarization angle $\beta_{F_{\rm max}}$ does not show any evolution, since, in the test case, the geometry of the system does not depend on time and the inclination of the binary orbit/scattering screen with respect to the line of sight is the only parameter determining the orientation of the polarization ellipse.
{\bf The periodic wobbling of the polarization angle with a frequency 
tracking (twice) that of the observed total flux is a unique feature of the MBHB scenario, and can be used to test such hypothesis for any variability selected MBHB candidate.}

\begin{figure*}
    \centering
    \includegraphics[width=0.47\textwidth]{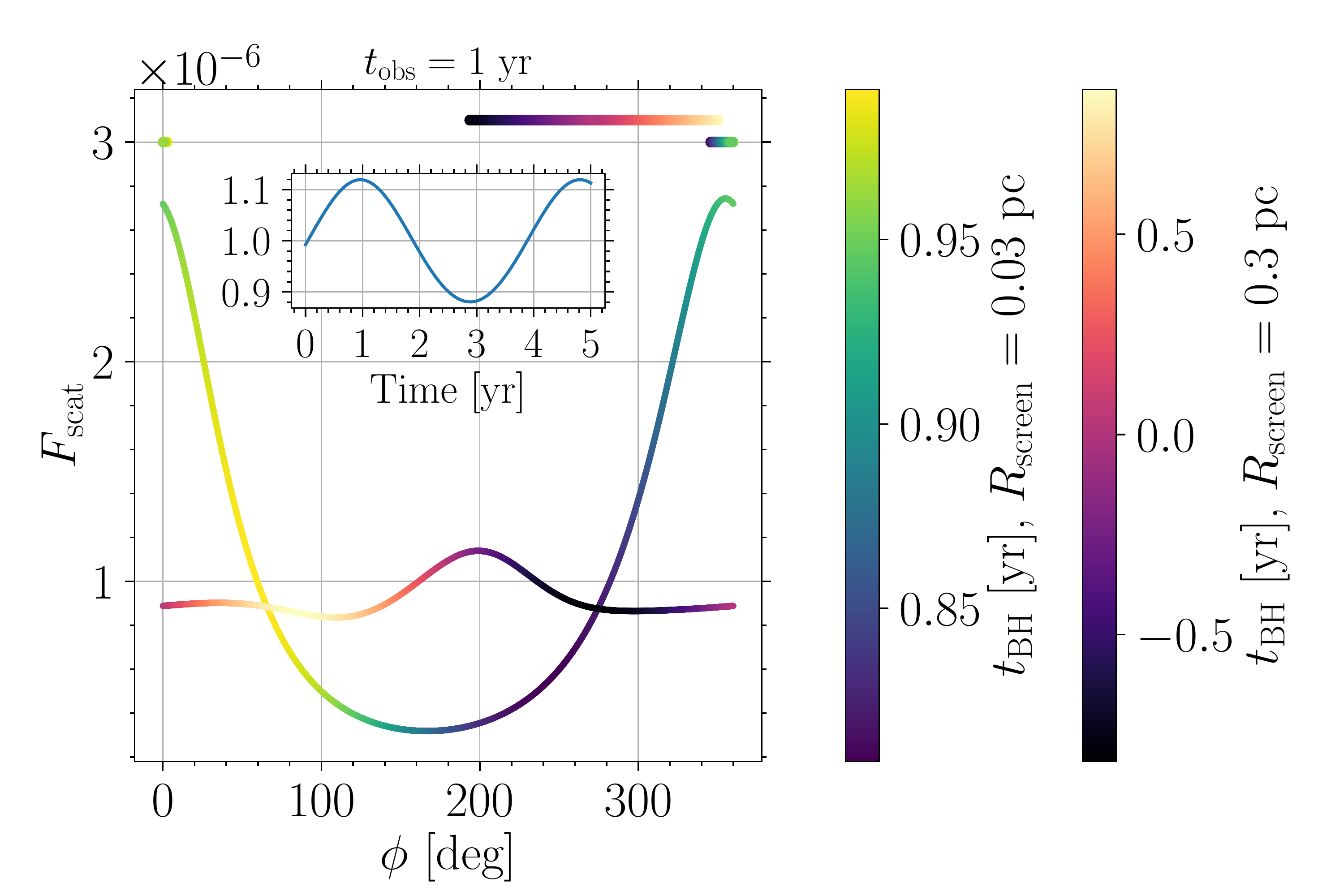}
    \includegraphics[width=0.47\textwidth]{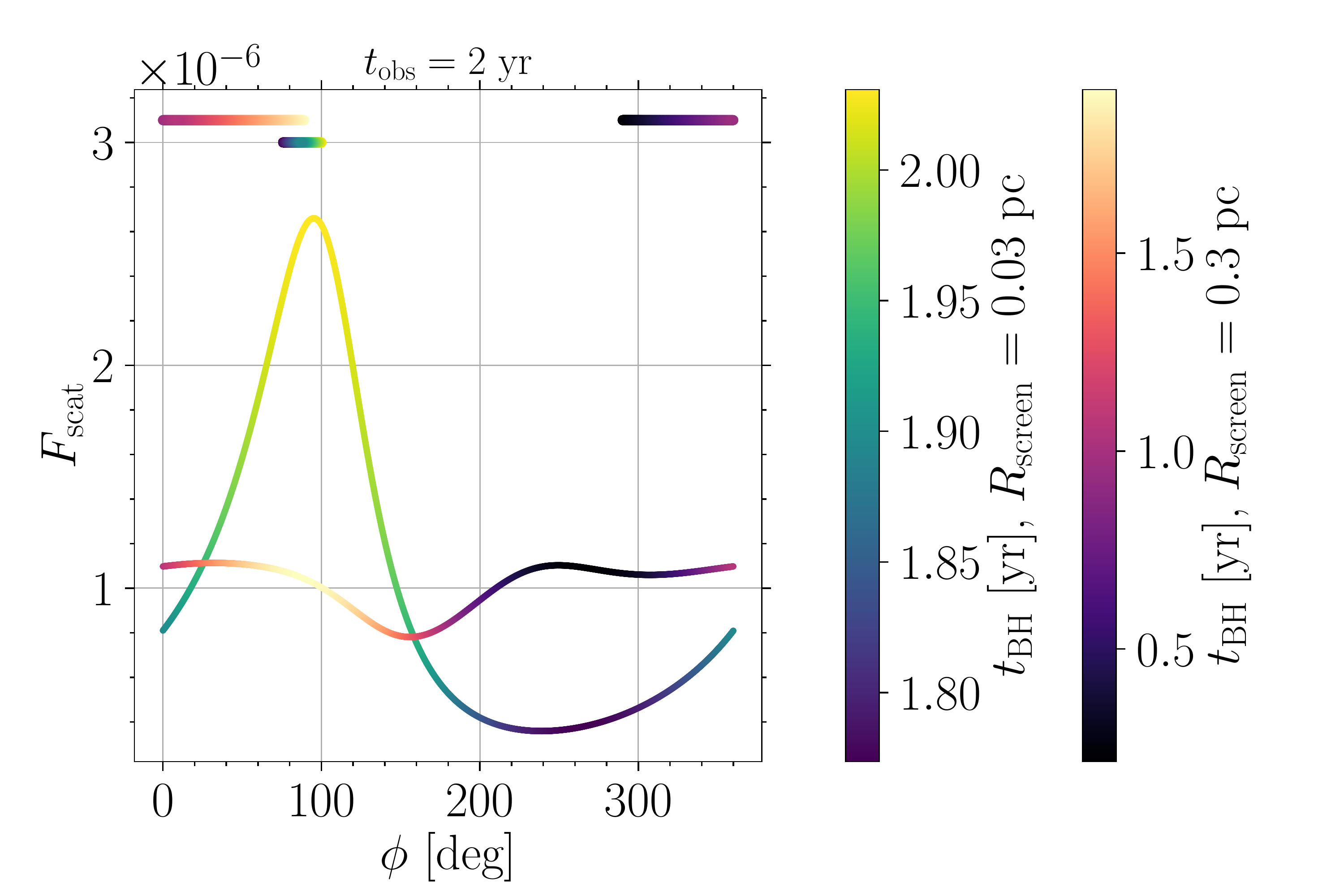}
    \includegraphics[width=0.47\textwidth]{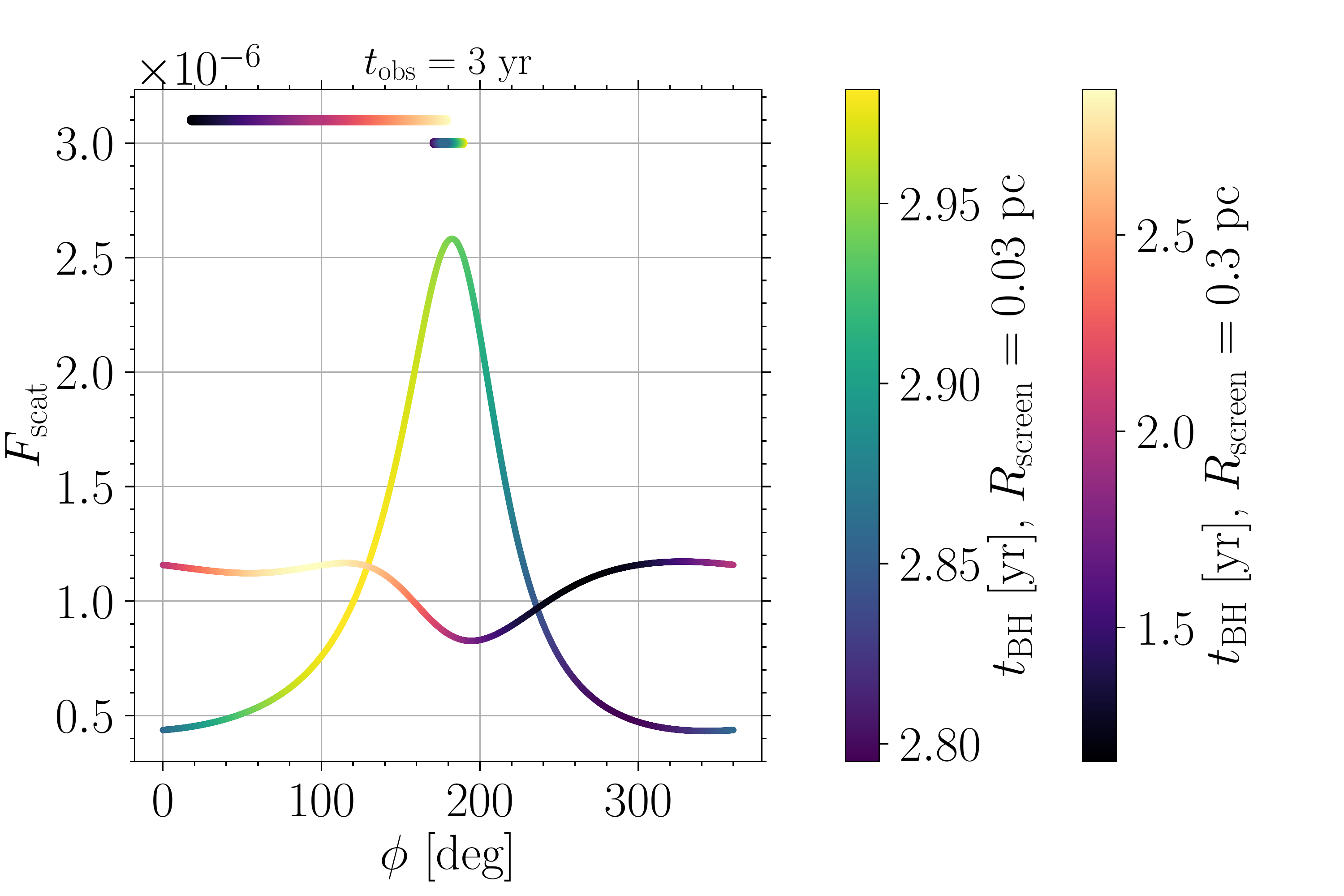}
    \includegraphics[width=0.47\textwidth]{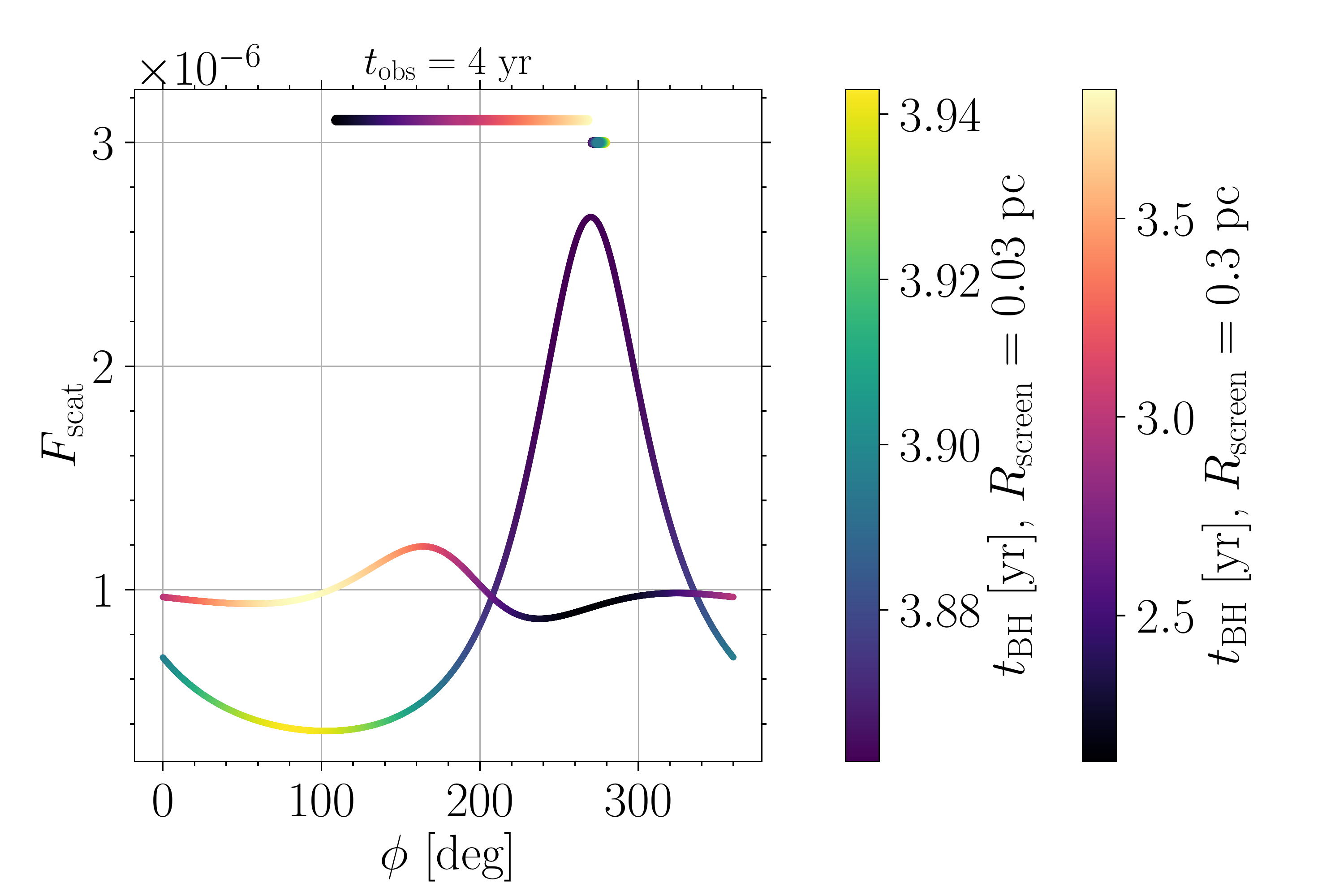}
    
    \caption{Relative contribution to the total scattered light $F_{\rm scat}$ of each scattering element, for the $R_{\rm screen}=0.03$ pc (blue-to-yellow color gradient associated with later times) and $R_{\rm screen}=0.3$ pc (black-to-white color scheme and earlier times). From top left to bottom right shown are four different moments corresponding to the time at which the observed polarisation is minimum, rising, maximum and declining (see time label at top of each plot). In all plots the color scale denotes the time at which the light was originally emitted by the secondary. The angular range spanned by secondary MBH during this time interval is marked near the top of each panel with a horizontal line using the same colour schemes used for the screen element contributions. For reference, we show the time evolution of the direct flux \Fdir in the inset in the upper left panel.}
    \label{fig:contributions}
\end{figure*}

Fig.~\ref{fig:contributions} quantifies the relative contribution of the different scattering elements to the total scattered light $F_{\rm scat}$ for the MBHB scenario at four different times, centred around the minimum and maximum P (upper and lower left panel, respectively), and in between (right panels). 
Since each screen element\footnote{Parametrized by the angle $\phi$ measured from the centre of the screen ring.} is characterized by a different time delay, we color code the time at which the radiation was originally emitted by the accretion disc of the secondary. The blue-to-yellow color gradient associated with later times refers to $R_{\rm screen}=0.03$ pc (i.e. to the case presented in Fig.~\ref{fig:res1}), while the other color scheme refers to a ten times larger screen radius, to clarify the effect of the delay times. All the other parameters are the same as the reference model shown in Fig.~\ref{fig:res1}. In the reference scenario the screen elements contributing the most at each time are those closer to the locii of the secondary orbit spanned during the time interval in consideration (highlighted in the figure as a horizontal line with the same color scheme). This is due to the short time delay between the direct and scattered light for $R_{\rm screen}=0.03$ pc ($\lsim 2$ months for every scattering element and secondary position, see eqs.~\ref{eq:t1}-\ref{eq:t2}). The contributions to the scattered light is instead more evenly distributed for a larger screen size (0.3 pc in the figure), the time interval at which such light has been originally emitted by the secondary is larger in this second case, and the screen region contributing the most is not the closest to the secondary anymore, as expected due to the larger time-delay between the direct and scattered light.  

\begin{figure*}
    \centering
    \includegraphics[width=0.47\textwidth]{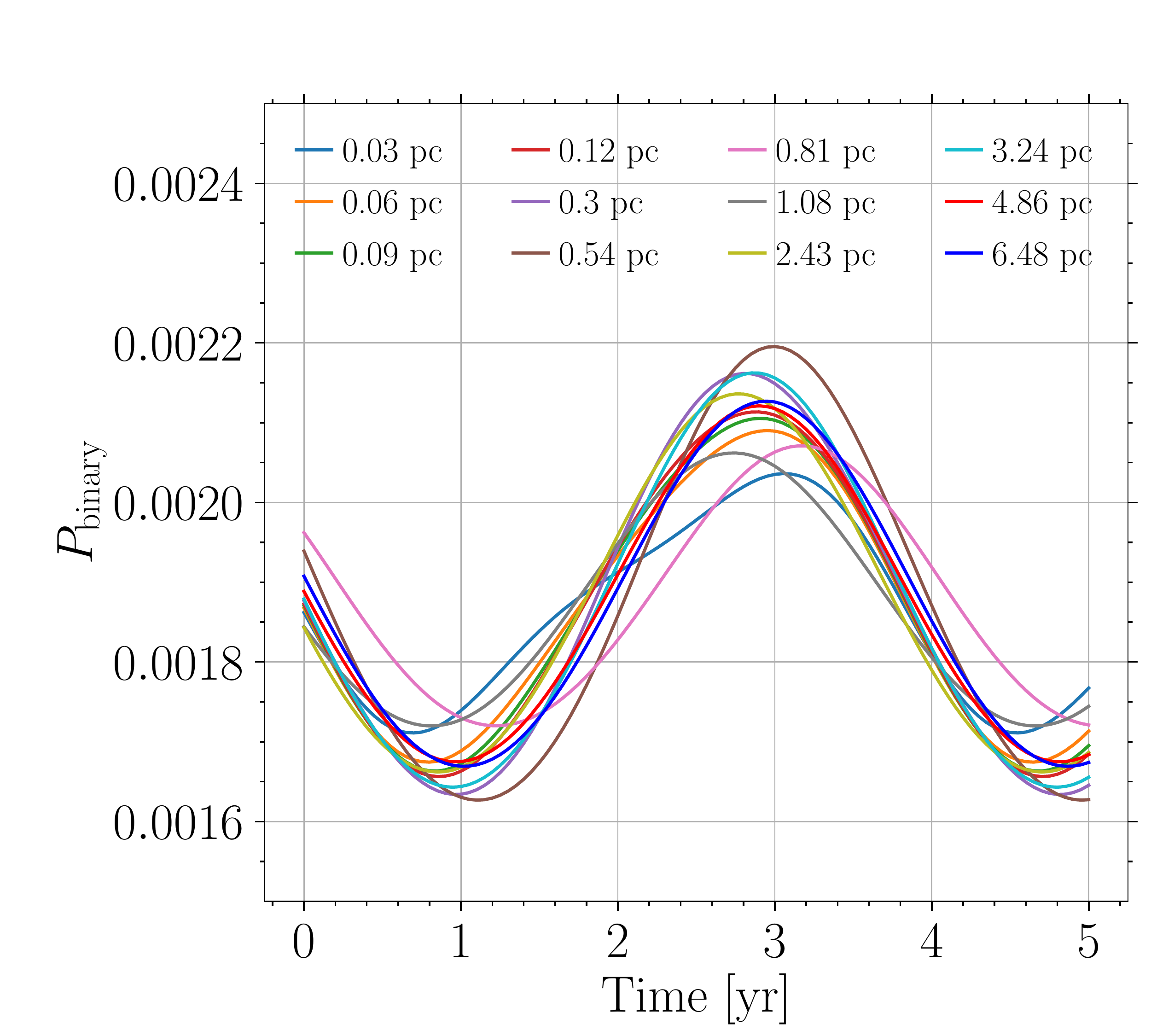}
    \includegraphics[width=0.47\textwidth]{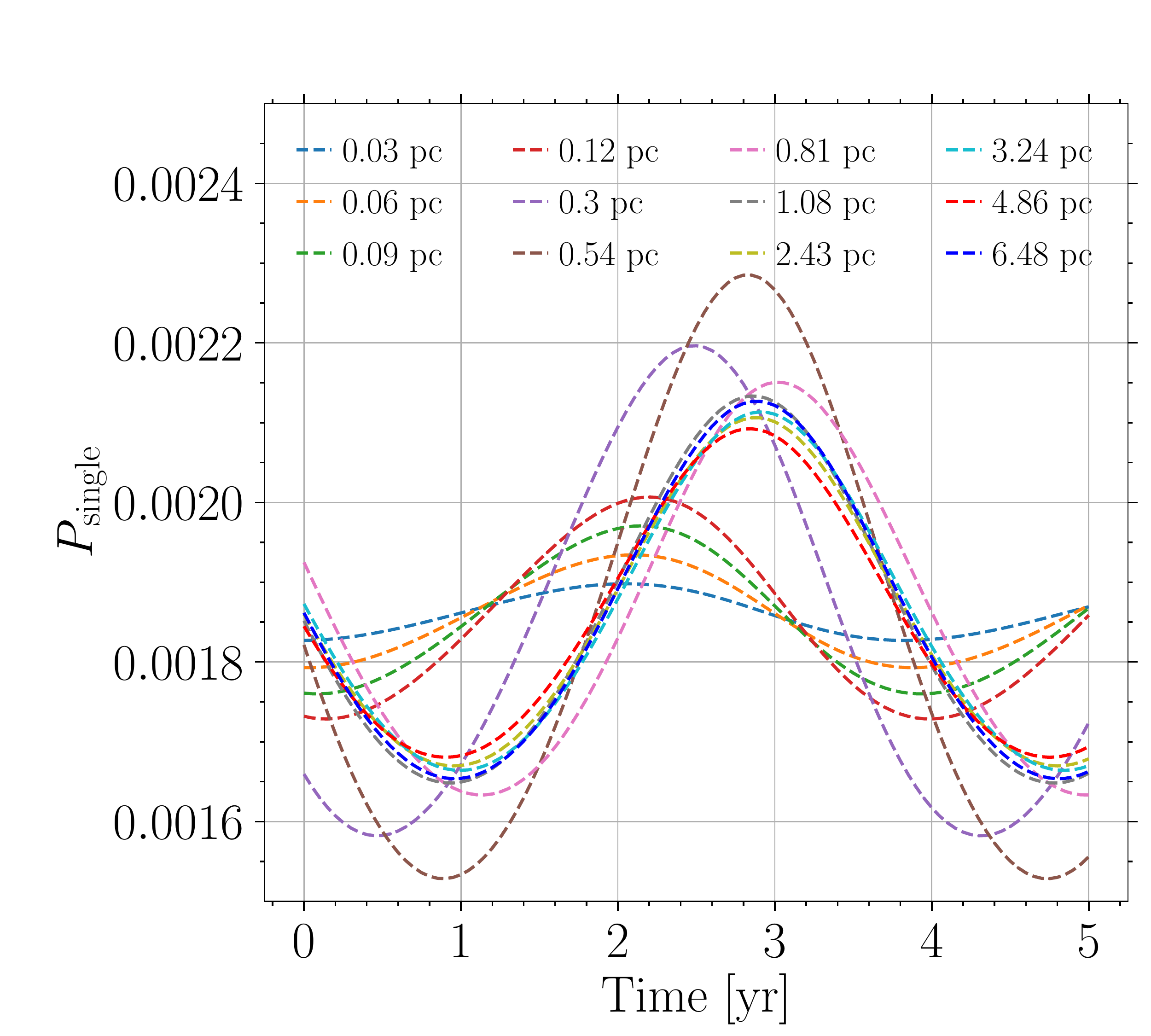}
    \includegraphics[width=0.47\textwidth]{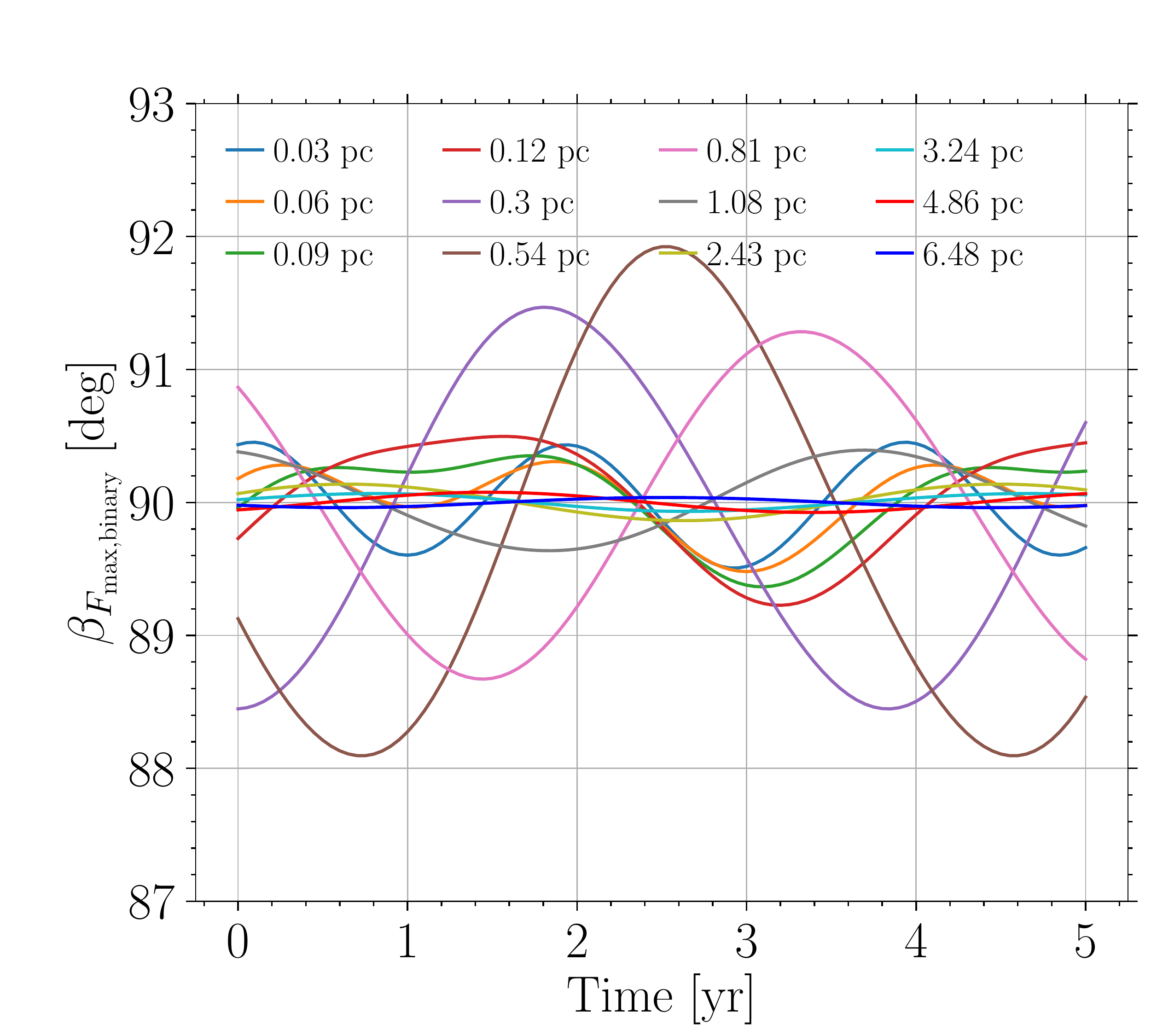}
    \includegraphics[width=0.47\textwidth]{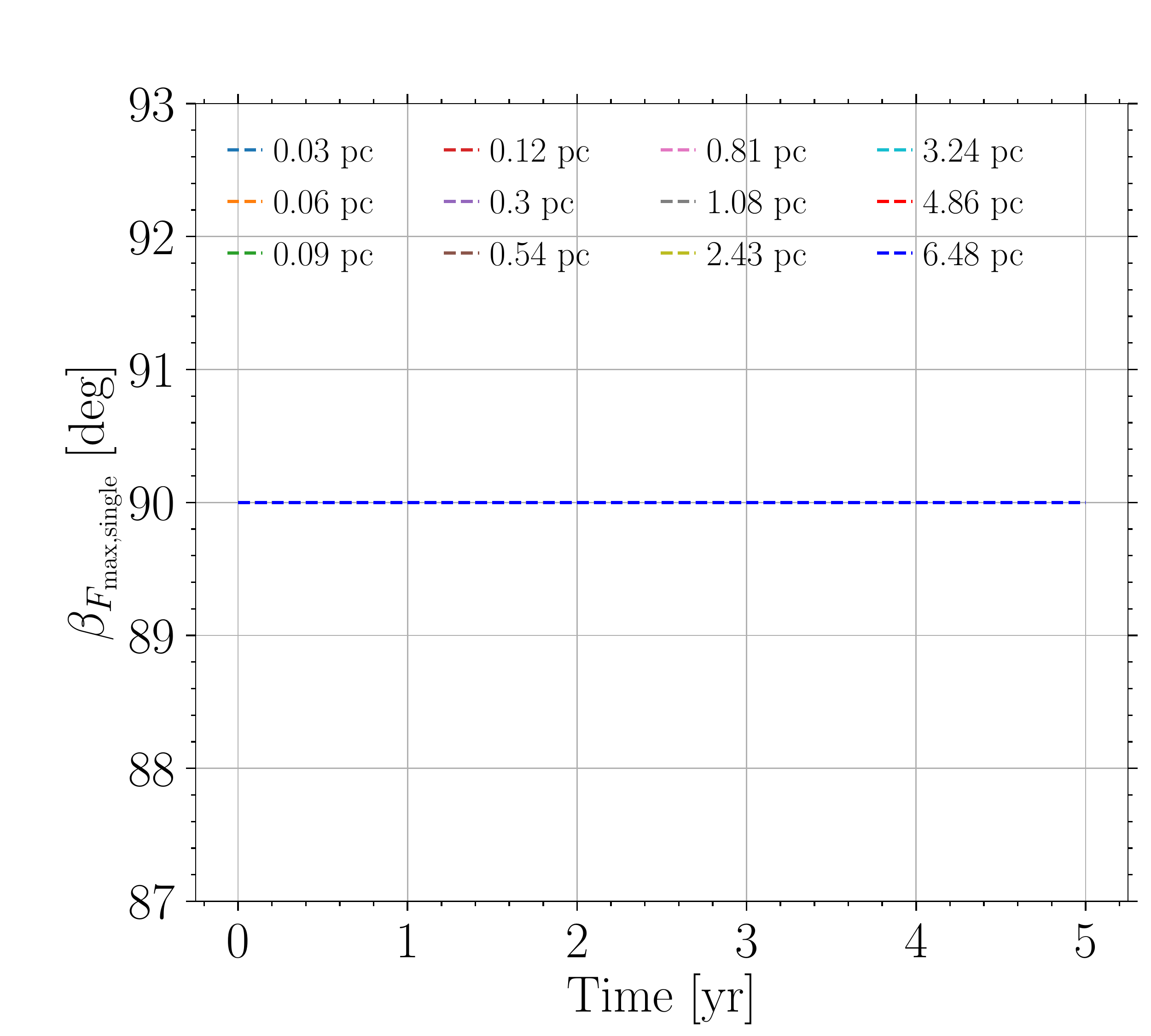}
   
    \caption{Top panels: Time evolution of the polarization fraction for the MBHB scenario (left panel) and for  the test case of a single MBH with an isotropically pulsating continuum (right panel) for different values of the scattering screen radius as labelled. Bottom panels: Time evolution of the polarisation angle for the same scenarios. Color code and line style as for the polarisation fraction. Note that in the single MBH scenario the angle does not vary.}
    \label{fig:sizes}
\end{figure*}

Fig.~\ref{fig:sizes} shows the effect of varying the scattering screen size on the polarization fraction and polarization angle for the binary case (left panels) and for the test case scenario with the single, isotropically pulsating MBH 
(right panels). The main difference in $P$ between the two models is in the time at which the maximum value is reached: the maximum of $P$ in the binary scenario is reached at about the minimum of the direct flux, as already commented above, regardless of the size of the scattering screen, while the peak for the single MBH scenario shows a stronger dependence on $R_{\rm screen}$, and stabilizes to the same value of the binary case only for values of $R_{\rm screen}$ significantly larger (by about a factor of 30) than the reference case. The main difference is again in the evolution of $\beta_{F_{\rm max}}$, showing an oscillating behaviour for the binary case while remaining constant in the single MBH case, regardless of the size of the scattering screen. Interestingly, the frequency of the oscillations around $\beta_{F_{\rm max}}=90^{\circ}$ evolves from twice the binary orbital frequency for $R_{\rm screen}\lsim 0.1$ pc $\lsim 6 \, a$ to the binary orbital frequency for larger $R_{\rm screen}$. The amplitude of the oscillations decreases for very large screen sizes ($R_{\rm screen}\gsim 1$ pc $\lsim 60 \, a$), with the binary model tending to the single case scenario for very large $R_{\rm screen}$, as it should.

\begin{figure*}
    \centering
    \includegraphics[width=\textwidth]{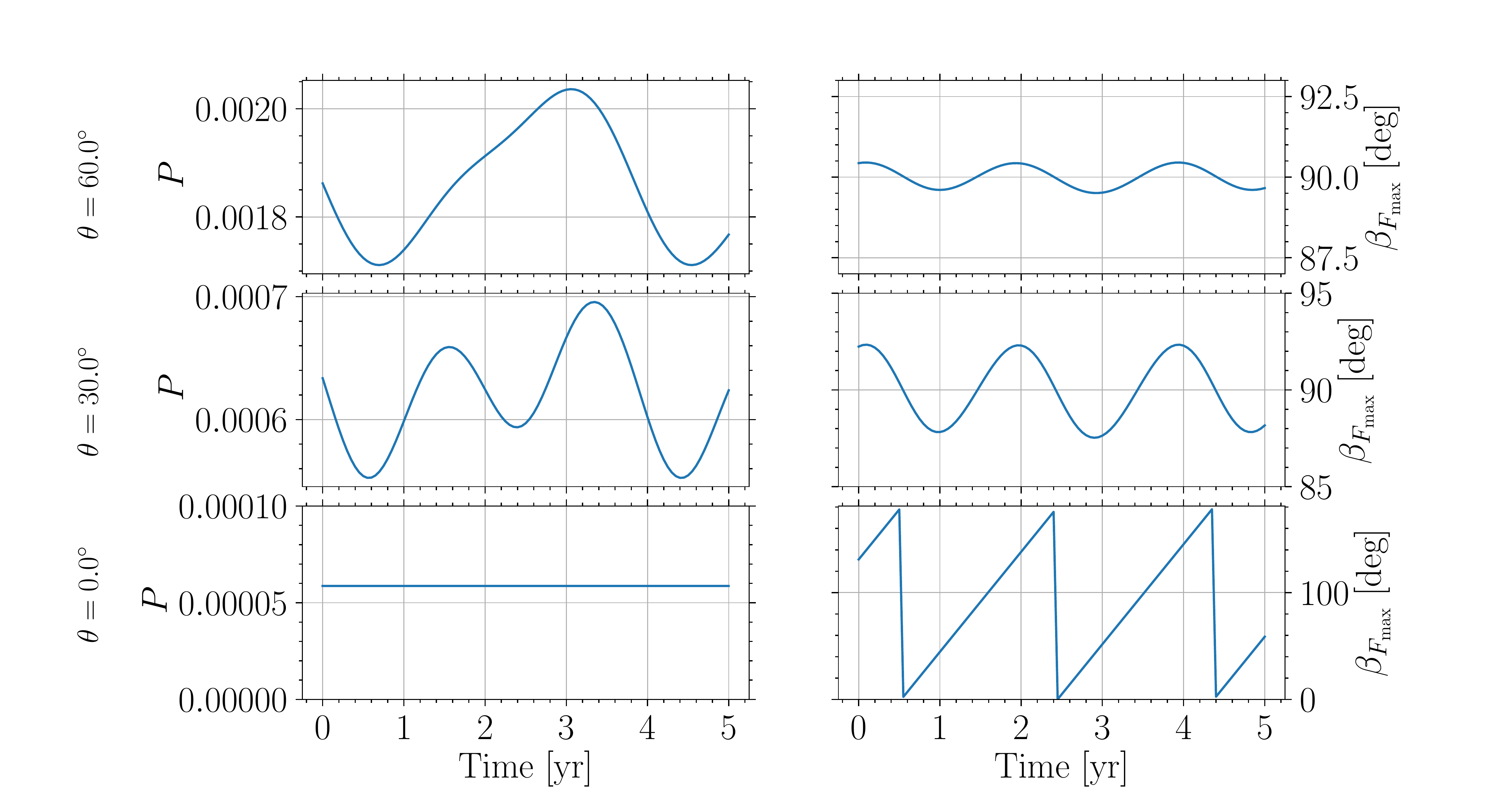}
    \caption{Left panels: time evolution of the polarization fraction $P$ for the binary scenario. The upper, middle and lower panels refer to relative inclinations between the MBHB angular momentum and the l.o.s. of $\theta = 60^{\circ}$, $30^{\circ}$, and $0^{\circ}$ respectively. Right panels: same as left panels but for the polarization angle $\beta_{F_{\rm max}}$. 
    Note that in the face-on case, the polarization angle rotates steadily on the sky over a full circle, at an angular frequency equal to twice the binary's orbital frequency.}
    \label{fig:resincl}
\end{figure*}

In Fig.~\ref{fig:resincl} we show how the relative inclination between the scattering screen (i.e. the MBHB orbital
plane) and the line of sight affects our results. The polarization fraction $P$ becomes smaller at decreasing inclinations, as expected for equatorial scattering in the single MBH scenario as well \citep{Smith05}.
Differently from the single MBH case, however, a residual polarization is still present in the $\theta=0$ (i.e. face-on) case here, due to the Doppler-boosting effect coupled with the varying relative distance between the secondary and the screen elements. 
The polar angle in this case steadily rotates on the sky over a full circle,
with a constant angular frequency equal to twice the MBHB orbital frequency.\footnote{Indeed, in the $\theta=0$ case the line of nodes is not defined, and the $x$ axis does not play any specific role.}
We stress however that such predictions for the small inclination cases will be hard to test, due to its overall low magnitude of $P$ and, most importantly, due to the negligible Doppler-boosting in the direct light for small inclinations, which would probably exclude such objects from any periodic AGN candidate sample in practice.

As a final test on the dependence of $P$ and $\beta_{F_{\rm max}}$ on the parameters of the system, we increased the $f$ parameter (the fraction of light scattered by the screen) by a factor of ten, finding the same behaviour observed in Fig.~\ref{fig:res1}, with the polarization fraction being 10 times higher, as expected as long as the total flux is dominated by the direct one.

We conclude examining the relative impact of the different physical prescriptions of the binary model, by comparing the standard model shown in Fig.~\ref{fig:res1} with four models with different implementations: $(i)$ a copy of the standard model not including any time delay between the direct and scattered light (dubbed `no delay'); $(ii)$ a standard model variation without any Doppler-boosting (`no boost'); $(iii)$ the same as scenario $(ii)$, but assuming a secondary emitting an intrinsically modulated isotropic flux that mimics the direct flux in the standard scenario (`no boost, modulated'); $(iv)$ the single MBH scenario with modulated direct flux (`single, modulated', previously shown with orange dashed lines in Fig.~\ref{fig:res1}). 

\begin{figure*}
    \centering 
    \includegraphics[width=0.48\textwidth]{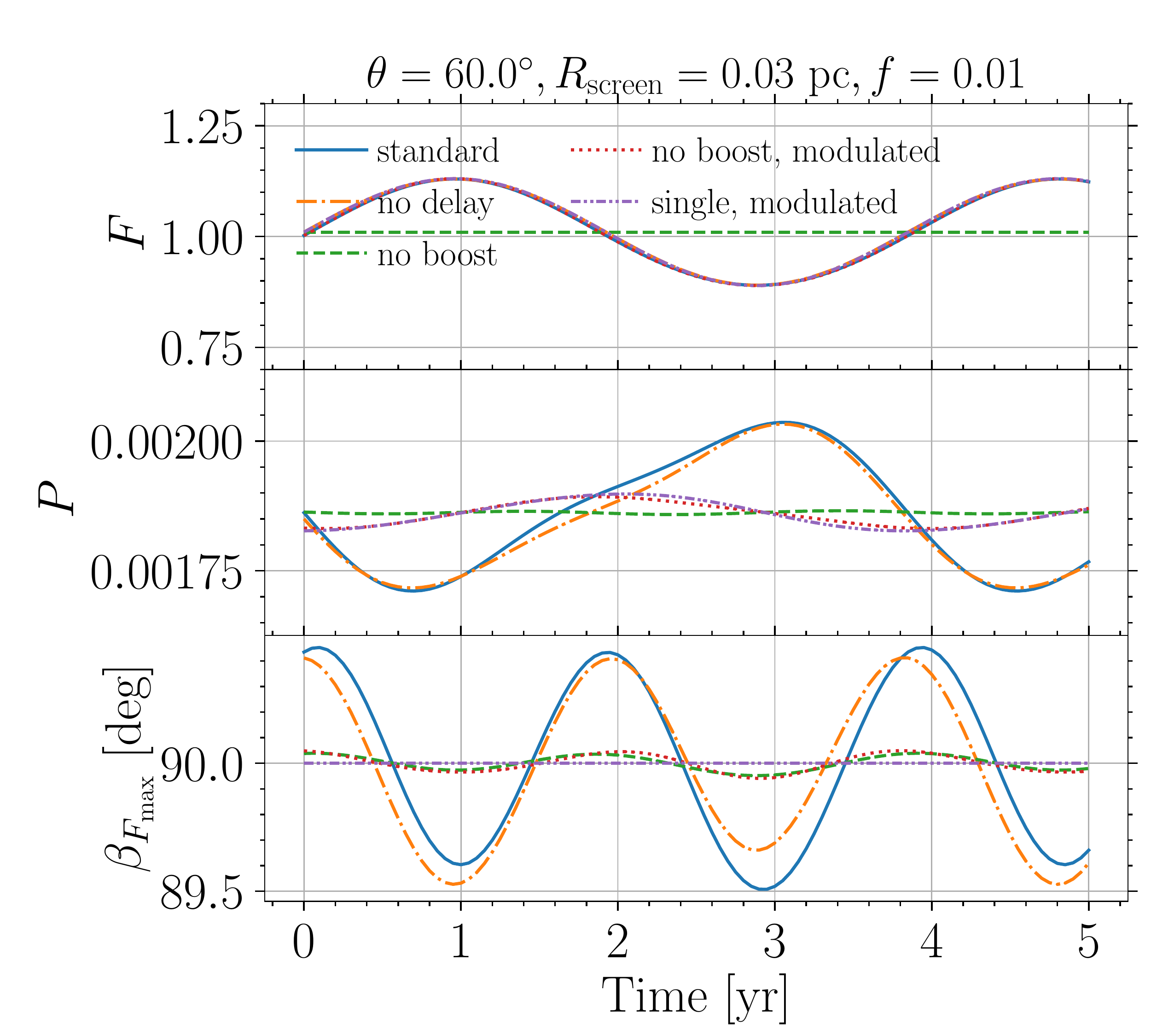}
    \includegraphics[width=0.48\textwidth]{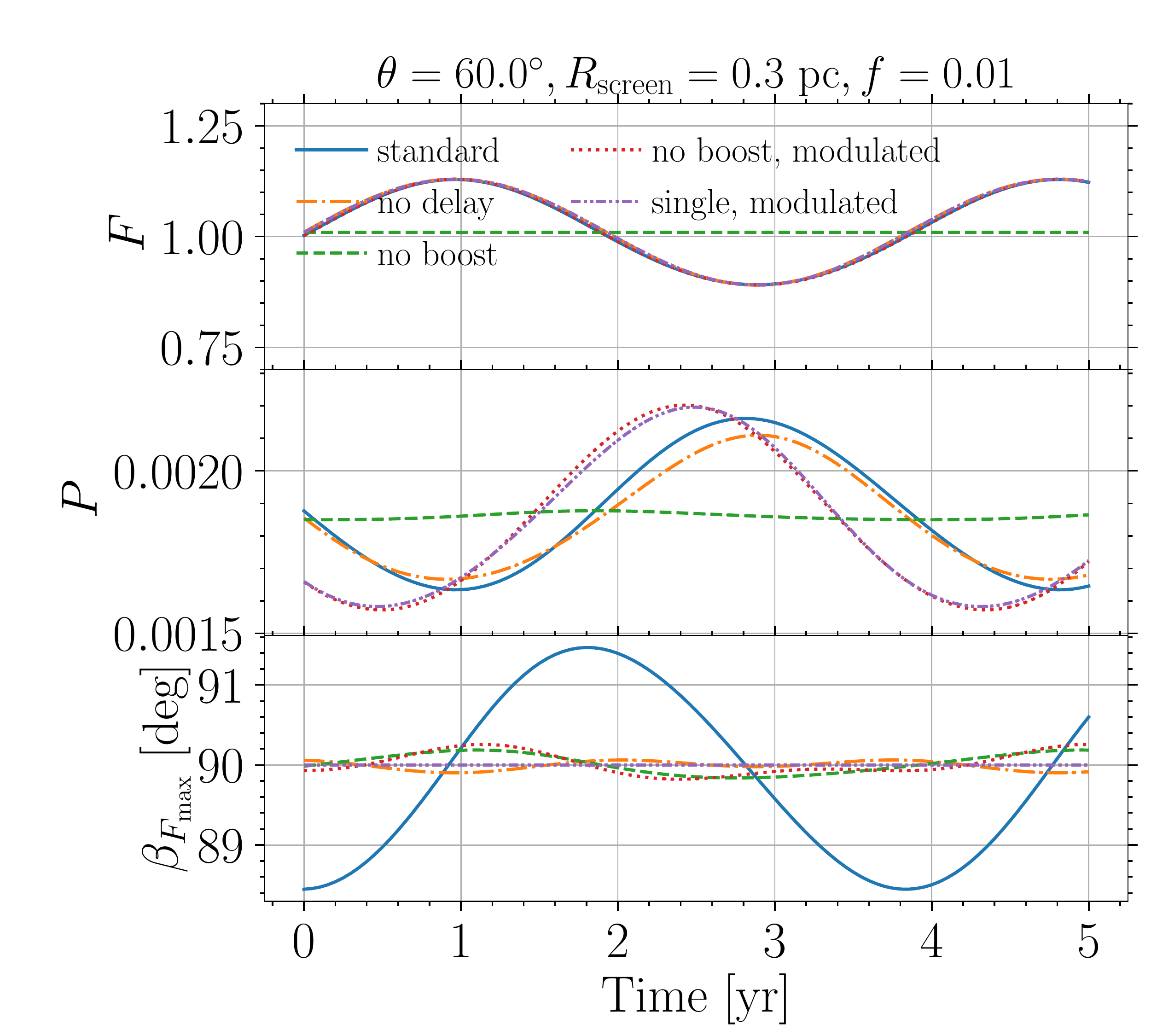}
    
    \caption{Impact of the different physical ingredients implemented in the model for $R_{\rm screen}=2a=0.03$ pc (left) and $R_{\rm screen}=0.3$ pc (right). The upper panels show the time evolution of the total flux, the middle panels that of the polarisation and the bottom panels show the evolution of the polarisation angle. The solid blue-line refers to the full model described in \S~\ref{sec:model} and shown in Fig.~\ref{fig:res1}. The alternative models without the inclusion of the boosting effect (dashed), without any time-delay between the direct and the scattered light (dashed dotted), without the boosting effect but with a modulated continuum (dotted) and that with only a single MBH at the centre of the screen emitting a modulated continuum are shown with different colours and linestyles (see labels).
  }
\label{fig:checkall}
\end{figure*}

The results of the above analysis are shown in Fig.~\ref{fig:checkall} for $R_{\rm screen}=0.03$ pc (left panel) and $R_{\rm screen}=0.3$ pc (right panel).\footnote{Note that the limits of the $y$-axis in the middle and lower panels are different from those in Fig.~\ref{fig:res1}, to better highlight the variations in the three observables.}
The observed total flux $F_{\rm tot}$ is similar for all the explored cases regardless of the assumed size of the scattering screen, except for the `no boost' case, in which $F_{\rm dir}$ is not modulated. The small variations observable in the other models are due to the different prescriptions on the sub-dominant component $F_{\rm scat}$ and result in significantly larger variations in the polarization fraction $P$ (middle panels) and polarization angle $\beta_{F_{\rm max}}$ (lower panels).
For the smaller ``reference'' size of the scattering screen the dominant contribution to $P$ and $\beta_{F_{\rm max}}$ is the relativistic boost of the secondary radiation, with the implementation of the time-delays playing a secondary role due to the short additional light path covered by the scattered light.
The same is true for $P$ for the larger screen (right panel), where a large modulation of the light impacting on the different screen element (i.e. larger than the variation obtained considering only the evolution of the relative distance between the secondary and each screen element) is needed to observe a variation of $P$ larger than $\sim 20\%$. The same order of magnitude in the variation of $P$ can be obtained by assuming an intrinsically modulated luminosity of the emitting MBH (as in the `single, modulated' and `no boost, modulated' models), but the predictions on $\beta_{F_{\rm max}}$ of these last two scenarios differ considerably with respect to those of the `standard' scenario: the oscillations of the polarization angle indeed can span up to $> 1^{\circ}$ only for the binary case when both the time delays and relativistic boost are considered. In this last case the effect of the time-delay is observable in the frequency of the oscillations of $\beta_{F_{\rm max}}$, which become equal to the binary orbital frequency for sufficiently large scattering radii, as discussed when commenting Fig.~\ref{fig:sizes}.

\section{Discussion}
\label{sec:discussion}

We have predicted polarization fractions at the $\lsim 0.3 f$ level with variability amplitudes of $\sim 5 \%$ for our reference model. In this study we chose a fiducial value of the fraction of scattered light to be $f=0.01$, and we have shown that the polarization fractions predicted scale linearly with $f$ as long as it remains significantly smaller than 1. This is consistent with the polarization fractions observed in type I AGN, $P \sim 1\%$ \citep[e.g.][]{Berriman+1990, Marin:2014}. 


Hence, adopting $f=0.01$ requires polarization fractions measurements of $P\sim0.2\%$, within an accuracy of $\sigma_P \leq 0.05\%$. This first requires a bright source, so that $1\%$ of the direct flux is still detectable. For bright enough sources (the higher polarization fraction the better), the polarization fraction has been shown to be measurable to the required values and precision.

The PlanetPol instrument, while no longer in commission, was capable of measuring $P$ down to $10^{-6}$ with $1\%$ accuracy  \citep{HoughPlanetPol:2006}. Furthermore, \citet{SmithPols+2002} have carried out polarization fraction measurements of Seyfert I galaxies, detecting sub-percent polarization fractions with uncertainties as good as $\sigma_P = 0.01\%$.


\citep{Berriman+1990} presents a compilation of the optical polarization fraction and polarization angles of the PG quasars. The MBHB candidate host PG 1302-102 is contained in this compilation. Two measurements were taken by \citet{StockMoorePols:1984}:
$P=0.18 \pm0.15\%$  
$P_{\theta} = 26 \pm 24$ deg. (May 21, 1979);
$P=0.08 \pm0.18\%$  
$P_{\theta} = 55 \pm 67$ deg.  (April 12, 1980).
Because this study only aimed for $0.2\%$ uncertainties in the polarization fraction, variability could not be discerned for PG 1302-102. However, modern polarimetry, aiming for precise polarization fraction measurements at the $\sigma_P \leq 0.01\%$ level could allow a determination of the polarization fraction variability of a candidate such as PG 1302-102, assuming it is indeed powered by a binary MBH.

In addition to the Doppler-boost's effect on the polarized flux, gravitational lensing may also play a role.
Just as the obscured broad-line region can be discerned in the polarized flux spectrum of Type II AGN, it could be possible to uncover `self-lensed' continuum emission in the polarized flux spectrum of accreting MBHBs. In the case where an aligned circumbinary disc is the scattering screen, as envisioned here, the incident radiation on the scattering screen, emanating from the secondary, will be periodically lensed by the primary BH \citep{DoDi:2018, DoDi:2019, Hu+2020}. This represents another modulation of the scattered and polarized flux, and future work should discern the properties of this polarized self-lensing signature.

Finally, the calculations carried out here assumed an aligned prograde disc as the scattering screen. For a misaligned or retrograde disc, the relative velocity between scattering screen and emitter, as well as the secondary Doppler-boost of the emitted radiation relative to the observer, is altered. This may alter the polarized light curves computed here, and further study of such additional degrees of freedom is required.


\section{Conclusions}\label{sec:conclusions}

In this study we focus on the Doppler-boosting model presented by \cite{DHS15} and \cite{DH17}, in which the phase dependent Doppler-shift of the (unpolarized) light emitted by the accretion disc of the secondary component of a MBHB results in the periodic evolution of the integrated in-band flux. We extend these models by predicting the time-evolution of the polarization properties of such in-band flux. The polarization is imprinted on the total observed flux by a scattered component, where the scattering elements have been assumed to have an equatorial geometry and lie on the MBHB orbital plane. We studied the dependence of the polarization features on the typical size of the scattering screen, on its inclination with respect to the line of sight, and on the fraction of scattered light.

We find that the total observed flux has {\bf (1) a clear and variable polarization}, with {\bf (2) an oscillating polarization angle on the sky}, centred around the direction of the semi-minor axis of the projected orbital ellipses (that is, perpendicular to the line of nodes).

Since the properties of the scattered light and, therefore, of the polarization features depend on the size and structure of the scattering screen, we cannot predict a typical shape of the time evolution of $P$. However, in all the cases we explored {\bf the polarization fraction always has a minimum in proximity of the observed maximum of the direct (and total) flux}. Such clear prediction, together with the characteristic evolution of the polarization angle, can be used as an independent confirmation of the MBHB nature of the AGN showing a varying light curve. Such selected candidates could then be followed-up through spectro-polarimetric observations in order to constrain the scattering screen geometries, searching for other specific signatures of the presence of MBHBs \citep{SMP19}, and allowing for a refined modelling of the polarization evolution of the continuum. The simple test presented here can therefore be used to compile a catalogue of electromagnetically selected MBHBs, necessary to inform current and future GW searches or to compare and cross-check with future GW-selected samples of MBHBs.

We conclude by enumerating the potential advantages of observing a MBHB candidate selected because of its modulated light curve, confirmed by a polarimetric follow-up, and lying within the sky localization error box of a future GW-detection through pulsar timing. 
Since this technique is sensible to near ($z\lesssim 1$) very massive objects ($M_{1+2} \gtrsim 10^8$ \msun) far from coalescence, the GW signal is monochromatic\footnote{Except some sources, when the pulsar term can be utilized \citep{CorbinCornish:2010}}. Due to the lack of frequency evolution\footnote{Observed in detections of stellar mass BHs made by ground-based interferometers and in the future observations of lighter MBH ($\sim 10^5$ \msun) using the space interferometer LISA.} the GW alone can directly measure neither the chirp mass $M_{\rm chirp}= (M_1 M_2)^{3/5} \, (M_1+M_2)^{-1/5}$ nor the luminosity distance $d_{\rm lum}$, but only an overall amplitude $A \propto M_{\rm chirp}^{5/3}/d_{\rm lum}$. Moreover, the measurement of $A$ is affected by large uncertainties due to its degeneracy with the MBHB sky localization, inclination and polarization angle, initial phase, all of which are not well determined by the GW detection alone \citep[see figure 7 in][]{SesanaVecchio2010}. This severely limits the amount of astrophysical information that can be extracted from a PTA signal.

An unambiguous identification of the host galaxy (e.g., in the Doppler-boost scenario, the detection of a periodically varying AGN with a frequency consistent with that of the  GW signal) associated to the GW detection would provide exquisite sky localization,\footnote{In the Doppler-boost scenario the initial orbital phase is constrained as well.}
decreasing the error on the GW amplitude. Most interestingly, the MBHB redshift can be determined, strongly constraining $d_{\rm lum}$ and, therefore breaking the degeneracy between it and $M\rm chirp$.

The electromagnetic signature proposed in this study provides additional precious information:
\begin{itemize}
    \item[$\bullet$] first, it constrains the polarization angle of the GW signal, determined by the orientation of the projected MBHB orbital ellipse on the plane of the sky, which we demonstrated can be constrained studying the polarimetric properties of the candidate. This allows for the complete description of the detector response pattern. In addition, the full orientation of the orbital plane, including its line-of-sight inclination as well as the orientation of the line of nodes, can be determined by fitting both by the Doppler boost of the direct signal and by the amplitude of the oscillations of $P$ and $\beta_{F_{\rm max}}$. Such 3-D information could be compared with the orientation a larger scale circumbinary disc/torus (possibly constrained with high-resolution imaging for sufficiently low-redshift systems), testing the occurrence of warps/misalignments in the gas distribution, likely driven by the binary potential itself~\citep[e.g.][]{Miller13};
    
    \item[$\bullet$] second, and perhaps more interestingly, the observed light curve provides $v_{2,\rm Z}$ and, given the constraints on $\theta$ both from the GWs and the EM signals, the magnitude of the secondary velocity $v_2\propto{M_1/\sqrt{M_{1+2}}}$ can be evaluated.\footnote{The semi-major axis is determined by the orbital frequency.} 
    For circular orbits such combination of the masses together with $M_{\rm chirp}$ is sufficient to measure both individual MBH masses. A similar procedure can be applied to a non-circular binary, for which the eccentricity can be constrained directly from the GW signal \citep{Taylor2016} and tested against the observed optical light curve.
\end{itemize}

While the polarization signatures we find here are subtle, and require percent-level measurements of polarization fractions as small as $O(1\%)$, we argued that such measurements should be within the capabilities of existing instruments, and could play a role in finding evidence for massive black hole binaries, and probe their characteristics.

\section*{Acknowledgments}
The Authors are grateful to Bruno Giacomazzo, Francesco Haardt, Alberto Mangiagli, Carmen Montuori, Albino Perego, Alberto Sesana and Mario Zannoni for the insightful discussions and suggestions.
MD and MB acknowledge funding from MIUR under the grant PRIN 2017-MB8AEZ.
DJD Acknowledges support from VILLUM FONDEN grant 29466.
ZH acknowledges support from NASA grant NNX15AB19G and NSF grants AST-2006176 and AST-1715661.
LH acknowledges support from the National Science Foundation of China (11721303, 11991052) and the National Key R\&D Program of China (2016YFA0400702).

\section*{Data Availability Statement}
The data underlying this article will be shared on reasonable request to the corresponding author.



\bibliographystyle{mnras}
\bibliography{biblio} 







\bsp	
\label{lastpage}
\end{document}